# Flexural free vibration as a non-destructive test for evaluation of viscoelastic properties of polymeric composites in bending direction

Mohammad Mehdi Jalili · Seyyed Yahya Mousavi · Amir Soheil Pirayeshfar

**Abstract** In many applications, viscoelastic properties of reinforced composites need to be determined prior to their real service life. Such properties can be assured by destructive and non-destructive tests. In this paper, a novel non-destructive test (NDT) method based on flexural free vibration is introduced to investigate the viscoelastic properties of fiber-reinforced composites. Three different types of industrial fibers (carbon, glass, and hemp) and an unsaturated polyester resin were selected to produce bar-shaped composites via pultrusion technique. These composite bars were used in a simple NDT method which just required a wooden hammer, two elastic jaws, a microphone and a recorder software program to perform the experiment. The composite bars were mounted on elastic jaws and hit by a wooden hammer at one end of the specimen as a perpendicular impulse. The produced sound at the other side of the bars was recorded. By analyzing the recorded sounds by means of fast Fourier transform (FFT), viscoelastic properties such as flexural elastic modulus and the loss parameter (Tan $\delta$) were calculated for the fiber-reinforced composites. For determination of flexural elastic modulus, the first three modes of vibration in FFT graphs were analyzed using Temuschenco theory. Also, Tan $\delta$ was determined by analyzing the drop in the intensity of vibrational energy as a function of time. Although there was a slight discrepancy between the calculated values from the NDT method and the actual data from dynamic mechanical thermal analysis (DMTA) approach, a good agreement was achieved between NDT and DMTA results.

## Introduction

Viscoelastic properties (such as elastic modulus and loss parameter, Tan $\delta$) are very important characteristics of polymeric composites and generally need to be determined before practical application of composite materials [1, 2]. To evaluate these properties, different test methods can be utilized for instantaneous destruction (e.g., DMTA) [3, 4] and non-destructive test (NDT) methods [4, 5]. One of the most important NDT methods is the resonance vibration involving mechanically vibrating a test specimen in a particular direction over a range of frequencies [6–8]. In the previous part of this research [9], a non-destructive test based on resonance-free vibration was introduced in which the specimens were vibrated in the longitudinal direction. This NDT method was employed to study the mechanical and dynamical properties of carbon fiber–polyester, glass fiber–polyester and hemp fiber–polyester composites in the longitudinal direction. The obtained results were also compared with those of dynamic mechanical thermal analysis (DMTA) approach indicating that there is a meaningful relation between the viscoelastic properties measured by the DMTA as well as the longitudinal free vibration non-destructive test [9, 10].

However, by the assistance of the method introduced in the previous article [9], only the longitudinal properties of materials could be determined and it was not applicable when the viscoelastic properties were required in the

M. M. Jalili (✉) · A. S. Pirayeshfar
Department of Polymer Engineering, Science and Research Branch, Islamic Azad University, Tehran, Iran
e-mail: m.jalili@srbiau.ac.ir

S. Y. Mousavi
Department of Polymer Engineering, Tehran South Branch, Islamic Azad University, Tehran, Iran

flexural direction. Knowledge of the flexural viscoelastic properties of a material is one of the main goals that engineers are seeking to acquire for many industrial and academic applications. In the article presented, which is a continuation of our previous work, a novel NDT method based on flexural free vibration is introduced to study the viscoelastic properties of carbon fiber–, glass fiber– and hemp fiber–polyester composites in the flexural direction. Further, we compare the flexural NDT results with those obtained from DMTA.

It should be noted that the theoretical and mathematical foundations of the longitudinal and flexural free vibration NDT are completely distinct due to their different test procedures.

**Experimental**

Sample preparation

To prepare fiber-reinforced composites, pultrusion technique was used. Also, three different types of fiber were selected and separately used including carbon fiber T300, provided from Troyca Co., USA, glass fiber WR3, provided from Cam Elyaf Co., Turkey, and also industrial-grade hemp fiber which was purchased from Iran Kenaf Co., Iran. Moreover, an isophthalic unsaturated polyester resin (Boshpol 751129) was purchased from Bushehr Chemical Industries Co., Iran, and cured using 1 vol % methyl ethyl ketene peroxide as initiator and 0.9 vol % cobalt octoate as accelerator. All these three components (i.e., polyester resin, initiator, and accelerator) were perfectly mixed before using in the pultrusion machine where the mixed resin was undergoing polymerization. Approximately, 70 vol % of each fiber was employed to fabricate the reinforced composites.

Destructive method

A three-point bending DMTA model, Perkin-Elmer DMA 8000 (USA), was utilized to examine the viscoelastic properties of the fiber-reinforced composite specimens at a heating rate of 5 °C/min and a frequency of 1 Hz. DMTA measurements were carried out in an ambient gas atmosphere. The composite samples were prepared with dimensions of $10 \times 34 \times 0.8$ mm$^3$.

NDT method

An NDT method based on flexural free vibration was employed to calculate the viscoelastic properties of the prepared composite bars which were cut with dimensions of $0.65 \times 0.65 \times 30$ cm$^3$ before testing. To set up the experiment, the bar-shaped specimens were placed on two elastic jaws and were hit by a wooden hammer at the end of the specimen as a perpendicular impulse. Meanwhile, a microphone was also positioned at the same position on the other side of the sample. Then, fast Fourier transform (FFT) was utilized to analyze the vibrating sound response which was recorded by the aid of Audacity software as a wave-format file. To determine flexural elastic modulus, the first three modes of vibration in FFT graphs were analyzed using Temuschenco theory [11]. In flexural free vibration test, elastic modulus can be calculated according to the Temuschenco theory [11]. This method was proposed in 1989 by Bordonne [11] to obtain the elastic and shear moduli of materials as a fast and reliable approach. According to the Temuschenco theory [11], the specific elastic modulus (the ratio of elastic modulus to the specific density) can be obtained by a linear regression on the values of $a_k$ and $b_k$ parameters as expressed in Eq. (1). The $a_k$ and $b_k$ parameters can be calculated from the FFT for the frequency of k$^{th}$ vibration mode according to Eqs. (2) and (3) [12, 13]:

$$a_k = (E/\rho) - [E/(K \times G)]b_k \qquad (1)$$

$$a_k = [4\pi^2 L^2 f_k^2 (1 + \alpha F_{1k})]/(\alpha X_k) \qquad (2)$$

$$b_k = [4\pi^2 L^2 f_k^2 F_{2k}]/X_k \qquad (3)$$

where $E$ is the elastic modulus, $\rho$ is the density, $K$ is a shape factor (in this research it is equal to 0.833), $G$ is the shear modulus, $l$ is the specimen length, and $f_k$ is the frequency of the k$^{th}$ vibrational mode obtained from analyzing the fast Fourier transform.

Furthermore, $\alpha$ is determined as follows:

$$\alpha = I/AL^2 \qquad (4)$$

where $I$ is the moment of inertia, $A$ is the cross area and $l$ is the specimen length.

$X_k$, $F_{1k}$ and $F_{2k}$ in Eqs. (2) and (3) can be also obtained from the following equations:

$$X_k = m_k^4 \qquad (5)$$

$$F_{1k} = \Theta^2(m_k) + 6\,\Theta(m_k) \qquad (6)$$

$$F_{2k} = \Theta^2(m_k) - 2\Theta(m_k) \qquad (7)$$

$$m_k = (2k+1)\pi/2 \qquad (8)$$

$m_1 = 4.73, \; m_2 = 7.8532, \; m_3 = 10.9956, \ldots$

$$\Theta(m_k) = [m_k \tan(m_k)\tanh(m_k)]/[\tan(m_k) - \tanh(m_k)] \qquad (9)$$

Also, the loss parameter (Tan $\delta$) was found out by analyzing the reduction in the intensity of vibrational energy as a function of time as comprehensively described in our previous work [9].

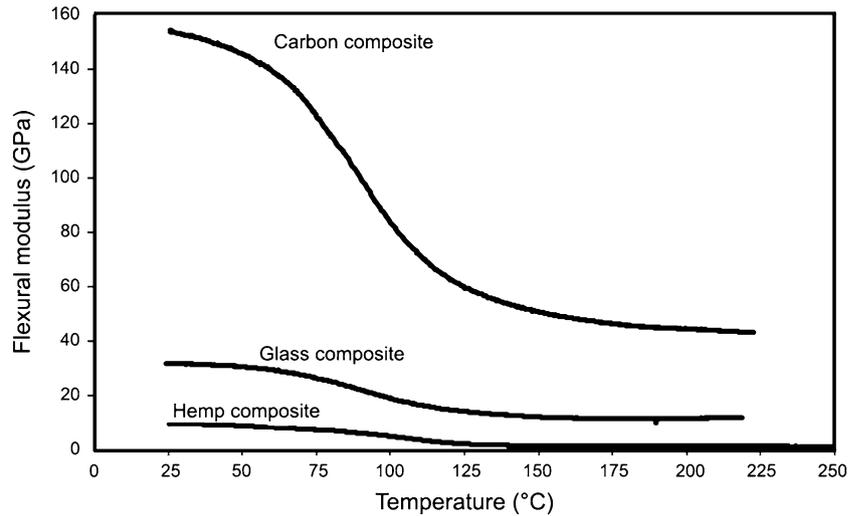

**Fig. 1** DMTA results for modulus versus temperature for carbon fiber–, glass fiber– and hemp fiber–polyester composites

In flexural free vibration NDT method, the first three resonance modes of vibration were evaluated, and each vibrational mode showed the loss parameter of that resonance mode [14]. The average of the three obtained Tan $\delta$ has been reported as the resulting loss parameter.

# Results and discussion

Figure 1 shows the behavior of elastic modulus obtained by DMTA approach as a function of temperature for all the composite specimens.

As shown, the flexural modulus decreases with increase in temperature for all the test samples. This phenomenon may be explained by softening of the resin due to the temperature increase. Since the purpose of this study is to compare DMTA and NDT test at ambient temperature, the flexural modulus of all specimens is selected from Fig. 1 at 25 °C and depicted in Fig. 2. The elastic flexural moduli for the fiber-reinforced composites resulting from non-destructive flexural free vibration test are presented in Fig. 2.

As evident in Fig. 2, the flexural moduli obtained from non-destructive flexural vibration test are greater than in DMTA data. By comparing these results, it is found that the moduli resulting from flexural free vibration method are approximately 23 % greater than the moduli resulting from the DMTA approach. This correlation is constant for all samples. In a previous research, this discrepancy was about 8–15 % where longitudinal free vibration NDT method was utilized [9]. It should be noted that in other studies, based on other NDT methods, acceptable differences between destructive and non-destructive methods have been also reported [15–17].

Regardless of the discrepancies, a good agreement is seen between the results measured through the DMTA

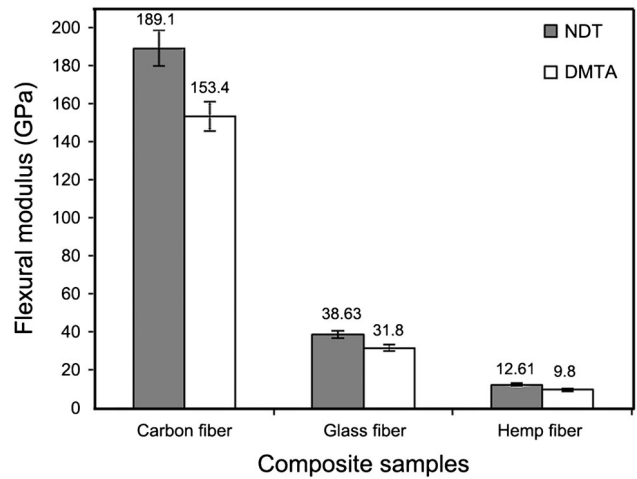

**Fig. 2** Flexural moduli obtained from NDT and DMTA methods at 25 °C for carbon fiber–, glass fiber– and hemp fiber–polyester composites

method and those calculated from flexural free vibration NDT. However, to introduce this NDT as a reliable method to determine the viscoelastic properties of polymeric composites, the interpretation of these differences is necessary.

To explain these differences, it should be considered that different methods cannot exactly result in identical values due to their basic theoretical assumptions and/or practical conditions. Moreover, it should be noted that after the specimens were tested using the NDT method, they were mechanically prepared for DMTA tests which might have affected some properties. The preparation process, involving some cutting and abrasion, introduces some stresses in the specimens and as a result lowers their performance. In view of this, the lower elastic modulus values resulting from the DMTA experiment can be explained.

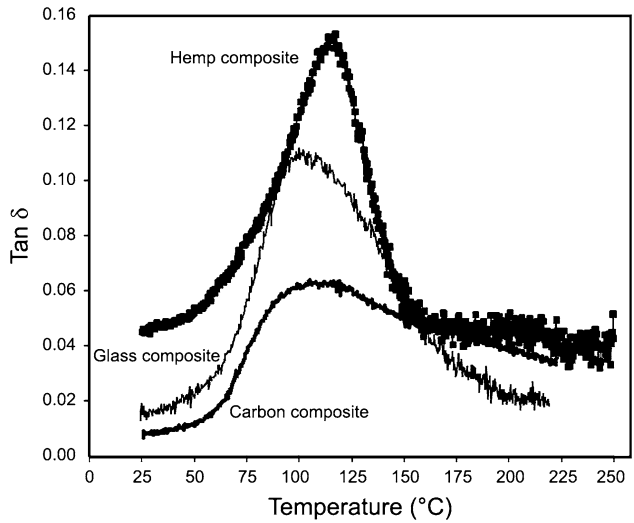

**Fig. 3** Tan δ graphs obtained by DMTA for carbon–, glass– and hemp–polyester composite specimens

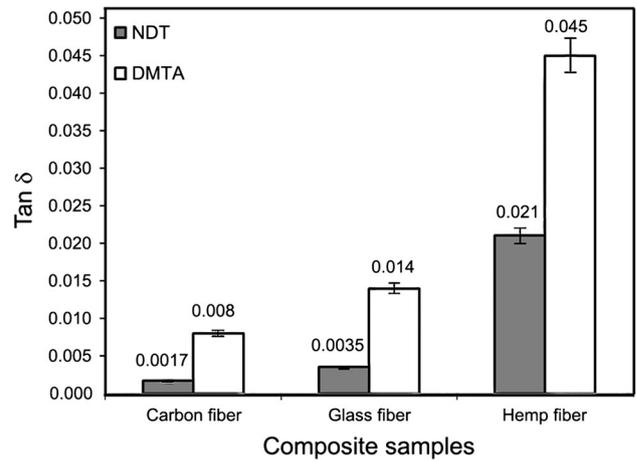

**Fig. 4** Loss parameters obtained from NDT and DMTA methods at 25 °C for carbon fiber–, glass fiber– and hemp fiber–polyester composites

One of the other reasons for such difference is the existence of the hinged support in the destructive flexural test. When the specimen is bending, some shear strain would be imposed besides the main bending strain. However, this shear strain is ignored and just the bending strain is considered as the total strain for such destructive flexural evaluation [18]. On the other hand, in non-destructive flexural free vibration no hinged support and as a result no shear strain are involved. This could be also considered as one of the reasons that explains the difference between the two tests of NDT and DMTA. Further, by applying bending forces, the appearance of the sample would change. As a result, the input forces would also change direction such that they would no longer be in the desirable initial direction. The other point that needs enough attention is that in the theory of destructive bending test, the area of the cross section is assumed to be constant, while in practice, this area decreases during the test process as the length increases in destructive bending tests. The above-mentioned points are the main reasons that could explain the difference between the elastic moduli obtained from the two DMTA and NDT tests [19].

Apart from that, polymeric composites are viscoelastic materials, so in these materials creep and phase angles occur in response to stresses [20, 21]. The term "viscous" in viscoelastic materials implies the length of time reacting to the forces, but the total time in which the NDT occurs is too short. In flexural free vibration NDT, when the sample is hit, the response sound would be recorded simultaneously, so the "viscous" does not completely affect the impulse due to the very quick hitting time. However, in DMTA test, since there is enough time for a "viscous" to react, it has its full effects. This behavior which has resulted from the time-dependent nature of polymeric materials could be also effective in producing different results.

Figure 3 shows the Tan δ graphs obtained by DMTA test for carbon fiber–, glass fiber– and hemp fiber–polyester composite specimens.

As it may be observed in Fig. 3, Tan δ increases as temperature rises because at higher temperatures, the polyester would become softer, so the elasticity of the resin would decrease. Since all NDT experiments have been performed at ambient temperature, to compare DMTA and NDT results, it is necessary to determine the value of Tan δ for all specimens at 25 °C. The obtained Tan δ (at 25 °C) from both DMTA and NDT tests for carbon fiber–, glass fiber– and hemp fiber–polyester composites are shown in Fig. 4.

According to Fig. 4, the behavior of Tan δ for all the three fiber-reinforced materials is the same in both NDT and DMTA methods. Also, the Tan δ values resulting from DMTA are greater than the values resulting from NDT. This indicates that composite samples show higher ability in dampening the energy during DMTA experiments. Also, the Tan δ values resulting from the flexural free vibration test are completely consistent with those obtained from the longitudinal free vibration test reported in our previous work [9]. However, as it can be seen in Fig. 4, the differences between the obtained results for Tan δ through NDT and DMTA are significant. It should be noted that the correction of the resonance vibration NDTs has been proved for many kinds of woods [13, 17]. So, it is essential to find the reasons for these differences in polymeric composites.

Such behavior may be related to the sample preparation procedure in which the composite samples were mechanically thinned to be fitted in the clamps of DMTA apparatus. This may have caused undesirable effects on the obtained loss parameters determined by DMTA. Furthermore, it could be also because of the slipping of the rigid composite samples on the DMTA clamps which was inevitable, especially for the polyester composites reinforced either by carbon or glass fiber. It is noteworthy that for the carbon fiber- and glass fiber-reinforced composites, the values of Tan δ obtained from the DMTA method are greater by two and three times than those obtained from the NDT method.

Evaluation of the samples and their results showed that the highly thinned composites prepared based on the DMTA device force limitation could have an influence on the DMTA results. Such significantly thinned samples, particularly those made from carbon and glass composites which could be easily bent even by hands, damped the input energies by their loose movements. So, due to this high damping ability, the values for Tan δ resulting from the DMTA can be expected to be higher than those obtained from the NDT.

It should be also mentioned that in Fig. 4, the differences in Tan δ resulting from NDT and DMTA seem to be very significant. This is because the values of Tan δ in the composite samples are very small, so any small difference looks very big regarding the initial values, while they are negligible without considering the primary values. On the other hand, the differences between loss parameters resulting from the NDT and the DMTA are small and acceptable.

In Fig. 5, loss parameters determined by DMTA and NDT methods are depicted in terms of their own modulus of elasticity for all specimens. As seen in Fig. 5, a relationship can be found between the elastic modulus and Tan δ values so that when the elastic modulus increases, the corresponding Tan δ decreases. It is noteworthy that this trend is exactly similar for both test methods, showing that the NDT results follow the behavior of DMTA, although there are some differences between them. The fact that NDT results follow the same pattern is another proof that the results are correct.

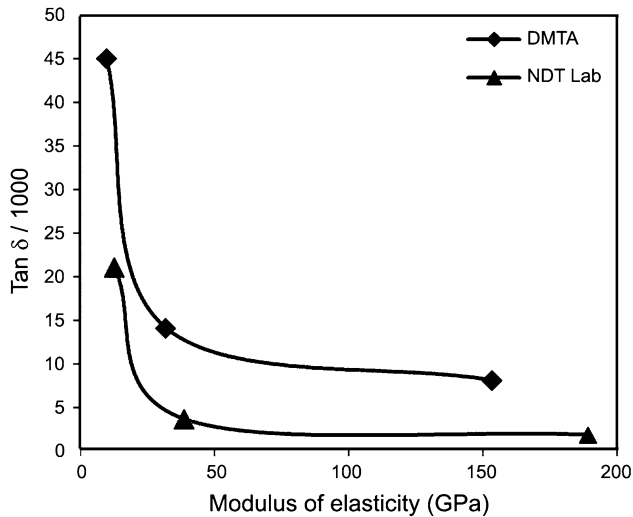

Fig. 5 Loss parameters resulting from both DMTA and NDT methods versus their own modulus of elasticity

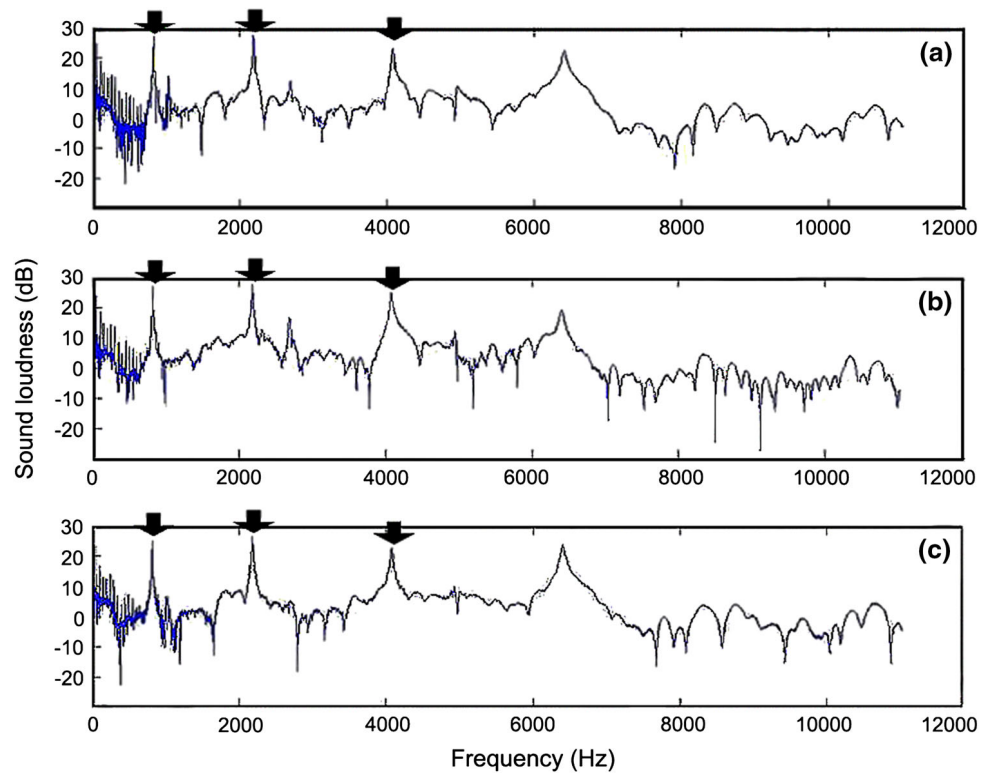

Fig. 6 The recorded sound using FFT technique for the carbon fiber–polyester composite specimen at its three replicates

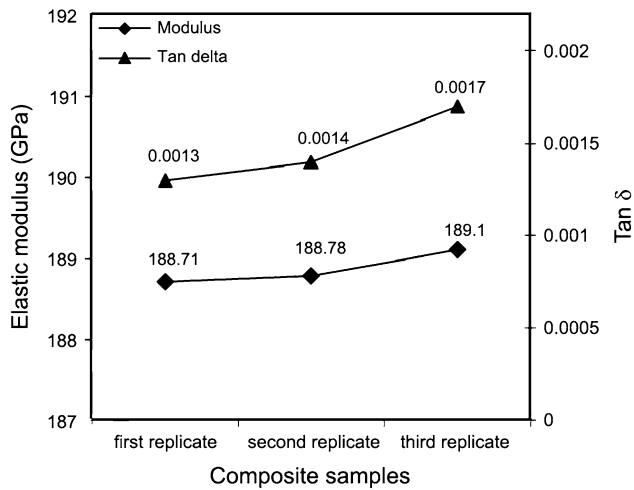

**Fig. 7** Elastic modulus and Tan $\delta$ plots calculated by means of flexural free vibration NDT from three separate replicates of carbon fiber–polyester composite specimen

Since the repeatability of a test method is an important factor to accept it as a reliable method, the carbon fiber–polyester composite was tested with three replicates using the introduced NDT method. Figure 6 shows the FFT of these replications.

The three graphs are similar and all the peak frequencies are almost the same. The moduli and the loss parameters obtained from analyzing these FFTs are depicted in Fig. 7.

As Fig. 7 shows, the obtained results from the three test replications for carbon fiber–polyester composite are almost identical and the differences are negligible. This shows that the obtained results are not random, indicating that the flexural free vibration NDT method is a repeatable approach. In fact, it is another proof that the results of the introduced NDT method are correct and accurate.

## Conclusion

In this study, the viscoelastic properties of carbon–, glass–, and hemp fiber–polyester composites in flexural direction were investigated using an NDT based on flexural free vibration. Comparing the results obtained by NDT with those measured from DMTA shows that the viscoelastic values calculated by NDT are acceptable and follow the same trend as DMTA results.

Given the reliable results of the NDT for analyzing the viscoelastic properties of polymeric composites and the easiness of implementing these methods, the more efficient introduction of these tests to polymer experts and running more research projects on other NDT methods are essential. In a future work, the authors would introduce another NDT method based on forcing the specimens to follow the vibrational frequencies of an electrical magnet.

**Acknowledgments** The authors are grateful to the Islamic Azad University, Science and Research Branch (IAUSRB), for financial support of the program.